\documentclass[reprint,aps,pra,superscriptaddress,notitlepage]{revtex4-1}
\usepackage{amssymb,amsmath}
\usepackage{graphicx}
\usepackage{dcolumn}
\usepackage{multirow}
\usepackage{color}
\usepackage{bm}
\usepackage[english]{babel}
\usepackage[caption=false]{subfig}
\newcommand{\red}[1]{\textcolor{black}{#1}}
\newcommand{\beginsupplement}{%
        \setcounter{table}{0}
        \renewcommand{\thetable}{S\arabic{table}}%
        \setcounter{figure}{0}
        \renewcommand{\thefigure}{S\arabic{figure}}%
        \setcounter{equation}{0}
        \renewcommand{\theequation}{S\arabic{equation}}
        \setcounter{section}{0}
        \renewcommand{\thesection}{S\Roman{section}}
        }

\begin{document}

\title{The critical role of substrate disorder in valley splitting in Si quantum wells}
\author{Samuel F. Neyens}
\author{Ryan H. Foote}
\author{Brandur Thorgrimsson}
\author{T. J. Knapp}
\author{Thomas McJunkin}
\affiliation{University of Wisconsin-Madison, Madison, WI 53706, USA}
\author{L. M. K. Vandersypen}
\affiliation{QuTech and the Kavli Institute of Nanoscience, Delft University of Technology, 2600 GA Delft, The Netherlands}
\author{Payam Amin}
\affiliation{Intel Corporation, Hillsboro, OR 97124, USA}
\author{Nicole K. Thomas}
\affiliation{Intel Corporation, Hillsboro, OR 97124, USA}
\author{James S. Clarke}
\affiliation{Intel Corporation, Hillsboro, OR 97124, USA}
\author{D. E. Savage}
\author{M. G. Lagally}
\author{Mark Friesen}
\author{S. N. Coppersmith}
\author{M. A. Eriksson}
\affiliation{University of Wisconsin-Madison, Madison, WI 53706, USA}

\begin{abstract}
Atomic-scale disorder at the top interface of a Si quantum well is known to suppress the valley splitting.
Such disorder may be inherited from the underlying substrate and relaxed buffer growth, but can also arise at the top quantum well interface due to the random SiGe alloy.
Here, we perform activation energy (transport) measurements in the quantum Hall regime to determine the source of the disorder affecting the valley splitting.
We consider three Si/SiGe heterostructures with nominally identical substrates but different barriers at the top of the quantum well, including two samples with pure-Ge interfaces.
For all three samples, we observe a surprisingly strong and universal dependence of the valley splitting on the electron density ($E_v$$\sim$$n^{2.7}$) over the entire experimental range ($E_v$=30-200~$\mu$eV).
We interpret these results via tight binding theory, arguing that the underlying valley physics is determined mainly by disorder arising from the substrate and relaxed buffer growth.
\end{abstract}

\maketitle
Gate-defined quantum dots in Si are attractive candidates for quantum bits (qubits), because of their weak spin-orbit coupling, natural abundance of nuclear-spin-zero $^{28}$Si, and compatibility with industrial scale fabrication techniques~\cite{Zwanenburg:2013p961}. 
However, Si qubits are affected by the conduction band valley degeneracy, which is twofold for devices formed in Si/SiGe quantum wells or at Si-MOS interfaces~\cite{Ando:1982p437,Schaffler:1997p1515}. 
The remaining degeneracy is lifted by a sharp quantum well interface.
The energy difference between these levels, known as the valley splitting, depends on details of the interface, including atomic-scale disorder, as well as the vertical electric field~\cite{Ando:1979p3089,Friesen:2006p202106,Friesen:2007p115318,Kharche:2007p092109}. 
For several types of silicon spin qubits, including single-spin~\cite{Veldhorst:2015p410, Yoneda:2018p102, Zajac:2018p439, Watson:2018p633}, singlet-triplet~\cite{Petta:2005p2180,Maune:2012p344,Wu:2014p11938,Nichol:2017p1}, and exchange-only~\cite{Gaudreau:2011p54,Medford:2013p654,Eng:2015p1500214}, the valley splitting should be large enough that only the lowest valley state is accessible during preparation, manipulation, and readout. 
Furthermore, valley splitting sets the energy scale for silicon-based quantum dot hybrid qubits~\cite{Kim:2014p70,Kim:2015p15004,Schoenfield:2017p64,Thorgrimsson:2017p32}, and should be in a range that is appropriate for AC gating ($\sim$10~GHz). 
For all qubit schemes, scalability will be enhanced when the valley splitting is as predictable and repeatable as possible.

Large valley splittings have been relatively easy to achieve in Si-MOS quantum dots, due to the combination of strong, tunable electric fields and abrupt SiO$_2$ interfaces~\cite{Yang:2013p3069,Gamble:2016p253101}, and in donor-based qubits it arises naturally from the strong three-dimensional confinement~\cite{Pla:2012p489,Harvey:2017p1029,Broome:2018p980}.
In Si/SiGe heterostructures, valley splittings tend to be smaller, making more important any variations in the valley splitting that can arise, for example, from variability in the sharpness and disorder of quantum well interfaces; experimental measurements reveal valley splittings ranging from tens to hundreds of $\mu$eV~\cite{Weitz:1996p542,Koester:1997p384,Lai:2006p161301,Goswami:2007p41,Shi:2011p233108,Shi:2014p3020,Mi:2015p035304}, 1-2 orders of magnitude below theoretical predictions for ideal quantum wells~\cite{Boykin:2004p115}.
Recent theoretical work predicts that specific alternating layers of pure Si and pure Ge at the quantum well top interface may significantly enhance the valley splitting~\cite{Zhang:2013p2396}.  
However, the added complexity could increase the atomic-scale disorder.
To minimize such effects, it is interesting to consider a simplified structure, reflecting the common element in each of the proposed heterostructures:  a thin, pure-Ge layer at the top of the quantum well.
As an added benefit, this structure has no alloy disorder in the active region, which could also suppress the valley splitting \cite{Kharche:2007p092109}.

Here we report the growth of heterostructures with a thin, pure-Ge layer at the top of the quantum well.  Structural characterization by scanning transmission electron microscopy (STEM) reveals this layer to be approximately 5 monolayers thick.  We report electronic transport measurements on three Hall bars, one each from two different heterostructures with such a thin Ge layer, and one from a conventional Si/SiGe heterostructure used as a control.  We find the electron mobility at a density of 4$\times$10$^{11}$cm$^{-2}$ in these samples is slightly lower in the presence of the Ge layer (56,000 and 70,000 cm$^2$/Vs, compared to 100,000 cm$^2$/Vs for the control sample).  Magnetotransport measurements performed on all three samples reveal well developed Shubnikov-de Haas oscillations and integer quantum Hall plateaus.  We report activation energy measurements for magnetic fields corresponding to filling factors $\nu$=3 and $\nu$=5, for electron densities ranging from 2.0$\times$10$^{11}$ to 5.5$\times$10$^{11}$~cm$^{-2}$.  These measurements reveal energy gaps, corresponding to the valley splitting, which vary from a minimum of 30~$\mu$eV up to 200~$\mu$eV, with the latter value attained for a sample with a pure Ge layer, at an electron density of $n$=5$\times$10$^{11}~$cm$^{-2}$ and filling factor $\nu$=3.  
While the relatively small differences in the measured mobilities and valley splittings between the studied samples at fixed electron density can be attributed to heterostructure modulations and the presence or absence of alloy disorder at the top of the quantum well, we observe a much stronger dependence of the valley splitting on the electron density and the corresponding vertical electric field, which is consistent across all three samples, including the control.
Tight binding theory, including both the experimentally applied electric field and interface disorder in the form of atomic steps, is able to explain this steep dependence on density. Based on the combination of these theoretical results and the experimental observations, we 
propose that disorder in the underlying substrate and relaxed buffer layer, which is nominally identical for all three samples, is a dominant contributor to the valley splitting and its dependence on electron density.

\begin{figure}[t]
\includegraphics[width=0.35\textwidth]{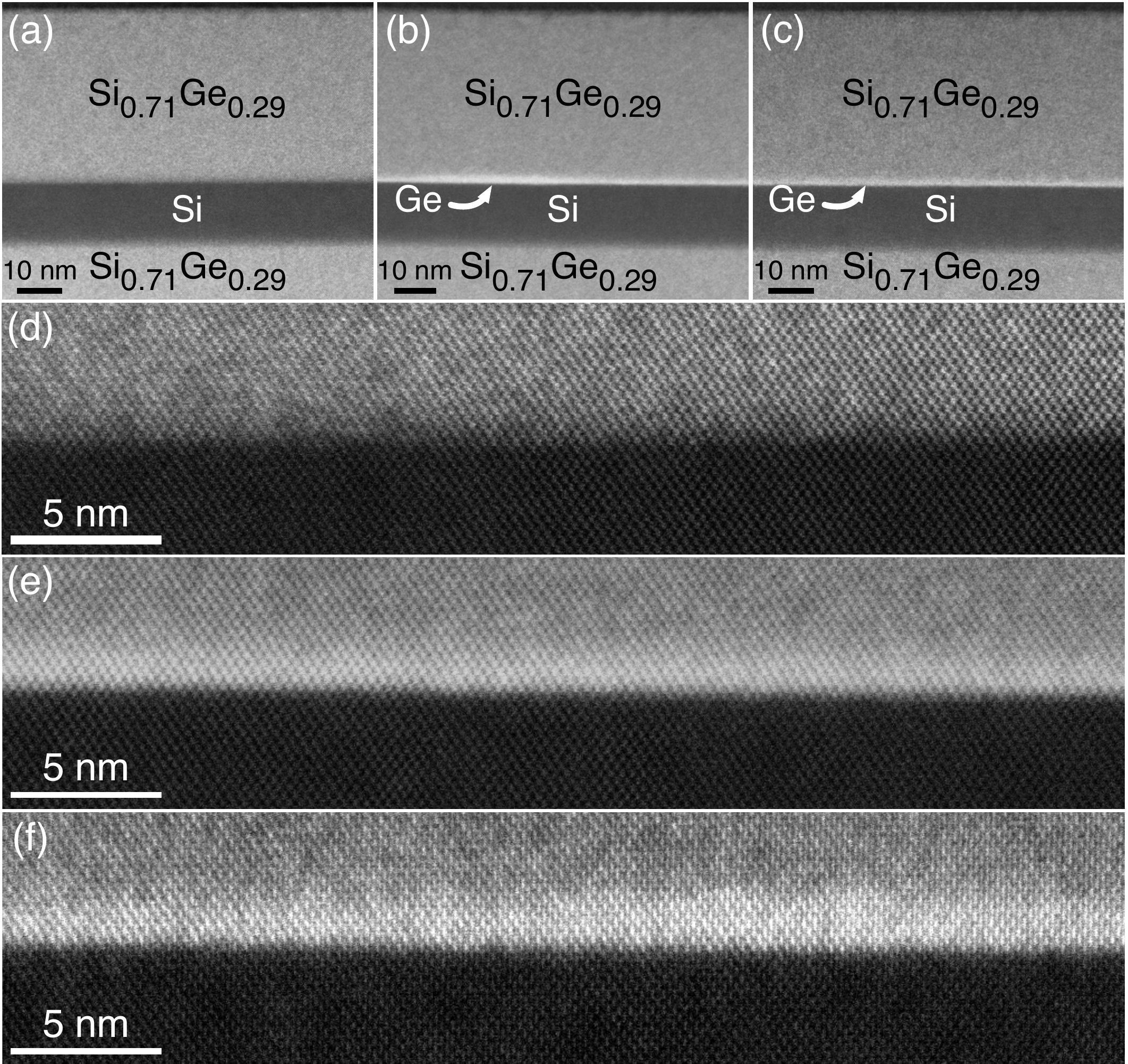}
\caption{\textbf{High-angle annular dark-field images of the three sample heterostructures, taken with a scanning transmission electron microscope (STEM).}
(a)-(c) Images of the quantum wells and barriers for samples A-C, respectively, taken directly below the accumulation gates of the Hall bars used to perform transport measurements. 
Brightness corresponds to Ge content, with Ge, SiGe, and Si appearing as white, gray, and black, respectively.
(d)-(f) High resolution images of the top quantum well barriers in (a)-(c).
}
\label{fig1}
\end{figure}

\red{All three samples are grown by CVD on commercially linearly-graded SiGe alloy with a final 2~$\mu$m Si$_{0.71}$Ge$_{0.29}$ layer that is chemo-mechanically polished.
Before the final CVD growth, these virtual substrates are ultrasonically degreased in acetone, then methanol, then rinsed in DI water.  
The native oxide is stripped in HF, DI rinsed, and then regrown in a UV-Ozone cleaner; this process is repeated once more.  
The samples are then Piranha cleaned, DI rinsed, and SC1 cleaned.  
After a final 5 minute DI rinse, the samples are dipped in 10\% HF for 20 seconds and loaded immediately into an LPCVD reactor where they are flash heated to 825$^\circ$C while silane and germane are flowing, then the temperature is reduced to the final 600$^\circ$C level.  
A 580~nm 29\% Ge alloy layer is deposited before growing the final well.}
For Sample A, the control, a conventional Si/SiGe heterostructure is grown. 
Samples B and C include a $\sim$1 nm thick interfacial layer of Ge above the Si quantum well. All three heterostructures have $\sim$13~nm Si quantum wells, followed by $\sim$34~nm barriers of Si$_{0.71}$Ge$_{0.29}$ (A) or Ge/Si$_{0.71}$Ge$_{0.29}$ (B,C), and $\sim$0.5~nm Si capping layers.  
The composition of each layer is set by the flow rates of the precursor gases: silane
for Si and germane
for Ge. For samples A and B, the growth is done continuously, at a constant temperature of 600$^{\circ}$C, ensuring that there is always active gas at the growth surface.
For sample C, at the top of the quantum well we lower the sample temperature to $<$400$^{\circ}$C to pause the growth while the reactive gas is changed from silane to germane, potentially yielding a more chemically abrupt interface with a modified disorder morphology.
We then exchange the Si and Ge precursors, while the sample is cold, and raise the temperature back to 600$^{\circ}$C to resume the growth of the Ge/Si$_{0.71}$Ge$_{0.29}$ barrier.

Figure 1 shows high-angle annular dark-field images of the three samples, taken with a scanning transmission electron microscope (STEM). 
The images confirm that samples B and C have a high concentration of Ge extending $\sim$1 nm above the Si quantum well, corresponding to $\sim$5~monolayers of material. 
The higher resolution images in Figs.~1(d)-1(f) suggest that all the samples have quite abrupt top quantum well interfaces; any differences in the abruptness are beyond the resolution of the STEM.

The undoped heterostructures were patterned with Hall bars of dimension 20 $\times$ 200 $\mu$m.
Ti/Au gates were evaporated on top of a 95 nm thick atomic layer deposition film of Al$_{2}$O$_{3}$, enabling \emph{in-situ} tuning of the electron density.
The mobilities of the samples at a density of 4$\times$10$^{11}$ cm$^{-2}$ are 100,000 for Sample A, 70,000 for Sample B, and 56,000 for Sample C, all in units of cm$^2$/Vs.  While the samples with Ge at the quantum well top interface have lower mobility, for all three samples the mobilities are consistent with previous demonstrations of quantum dot devices in Si/SiGe heterostructures~\cite{Simmons:2009p3234, Wang:2013p046801}.

\begin{figure}
\includegraphics[width=0.36\textwidth]{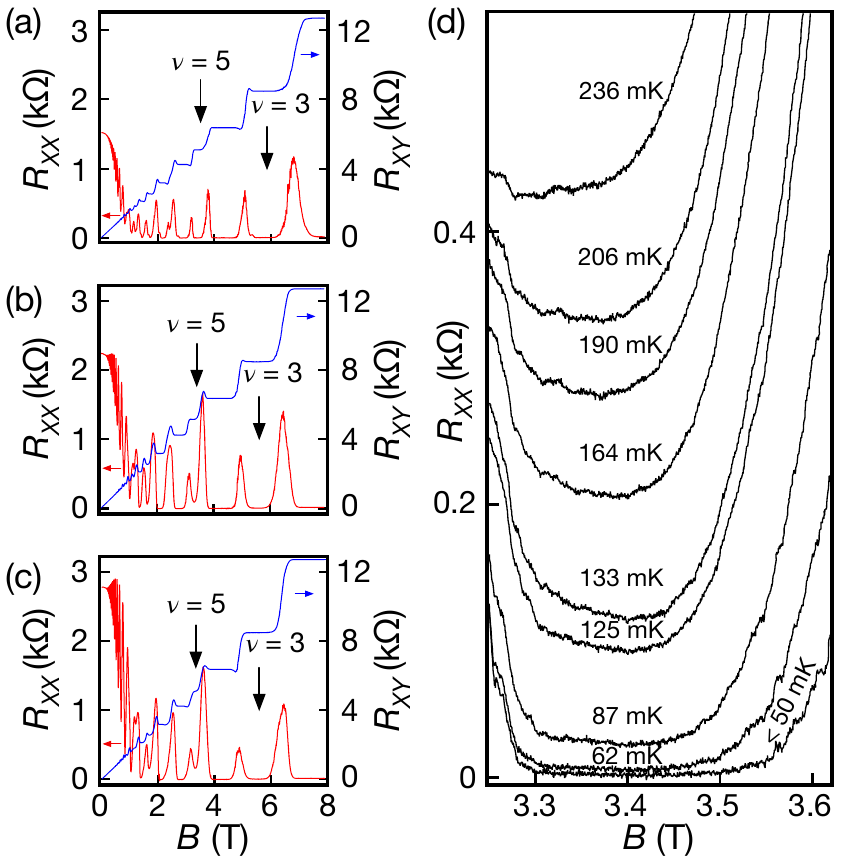}
\caption{ \textbf{Quantum Hall and thermal activation measurements}
(a)-(c) Longitudinal (red) and transverse (blue) resistances for samples A-C, respectively, as a function of magnetic field, acquired at base temperature. 
The $R_{XX}$ minima corresponding to valley splitting occur at odd-numbered filling factors ($\nu$). 
The $\nu$=3 and 5 minima, where we measure valley splitting, are indicated. 
(d) Activation measurements of sample A, at $\nu$=5. 
The mixing chamber temperature for a given $B$-field sweep is indicated above each curve. 
All measurements are taken at a carrier density of 4$\times$10$^{11}$ cm$^{-2}$.
}
\label{fig2} 
\end{figure}

In Fig.~\ref{fig2}, we report the magnetoresistance of all three samples in a cryostat at base temperature ($<$50 mK).
Shubnikov-de Haas minima in $R_{XX}$ occur when the Fermi level lies in the Landau level gaps
with odd-numbered filling factors ($\nu$), corresponding to valley splittings~\cite{Schaffler:1997p1515}. 
We measure the temperature dependence of $R_{XX}$ by fixing the magnetic field, heating the sample  to $\sim$250~mK, and allowing it to cool slowly while measuring $R_{XX}$.
A typical data set is shown in Fig.~2(d).
In the activated regime, the minima follow an Arrhenius scaling, $R_{XX} \propto e^{-E_{v}/2k_{B}T}$~\cite{Weitz:1996p542}, allowing us to determine the mobility gap $E_{v}$ corresponding to valley splitting (see supplementary material).
The primary source of uncertainty arises from the choice of the temperature range for the fitting. 
At lower temperatures, $R_{XX}$ is dominated by hopping conduction rather than activation, yielding a nonlinear Arrhenius plot~\cite{Ebert:1983p625}. 
A departure from linearity also occurs at high temperatures, as the $R_{XX}$ minima begin to shift in position and disappear~\cite{Usher:1990p1129}. 
To exclude these effects, we perform the fits over ranges that appear linear by eye on Arrhenius plots, and we estimate the uncertainty in the slope by varying the fitting range until it includes clearly nonlinear regions.

\begin{figure}
\includegraphics[width=0.34\textwidth]{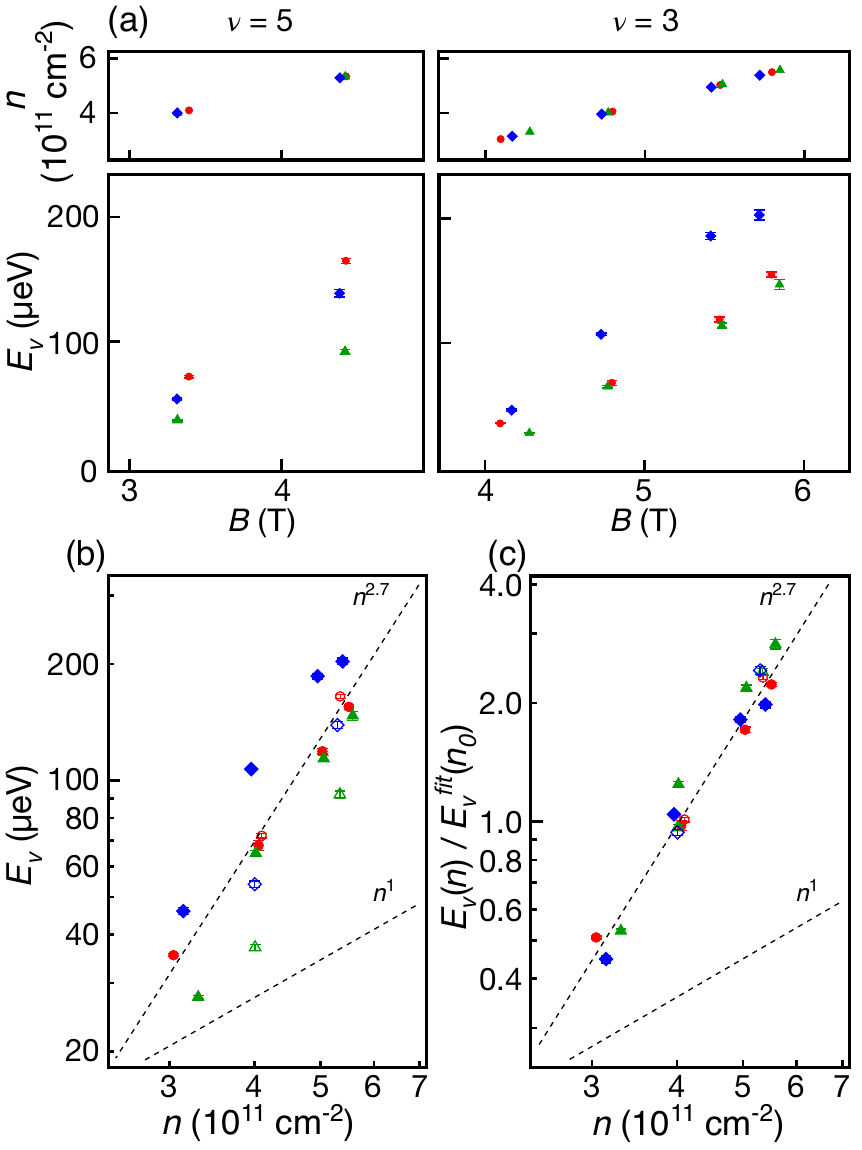}
\caption{\textbf{Valley splitting as a function of magnetic field, filling factor $\bm \nu$, and carrier density $\bm n$.} 
(a) Valley splitting in sample A (red circles), sample B (green triangles), and sample C (blue diamonds), at $\nu$=3 and 5, as a function of magnetic field. 
Here, we adjust $n$ so that $\nu$ remains fixed (top panels).  
(b) Valley splitting as a function of $n$, with $\nu$=3 (filled markers) and $\nu$=5 (open markers) plotted on the same graph. 
Dashed lines indicate linear and power-law functions of $n$, with $n^{2.7}$ yielding the best fit for all data sets.
(c) Scaled plot of the same data, to highlight the power-law scaling. Each data set is scaled by the fit value at $n_{0} $=4$\times$$10^{11}\!$~cm$^{-2}$. 
}
\label{fig3}
\end{figure}

As a control experiment, and to compare the mobility gap to an expected single particle gap, we apply this method to the Zeeman splitting of the $\nu$=6 Landau level.
We obtain a gap from the Arrhenius fits corresponding to Land\'{e} $g$-factors of $2.2\pm0.2$ for sample A, $1.8\pm0.1$ for sample B, and $1.9\pm0.2$ for sample C, close to the expected single particle value of $g$=2 for Si and providing an indication of the difference between the single particle gap and the mobility gap in these samples~\cite{Prange1990}.

\begin{figure}[t]
\includegraphics[width=0.28\textwidth]{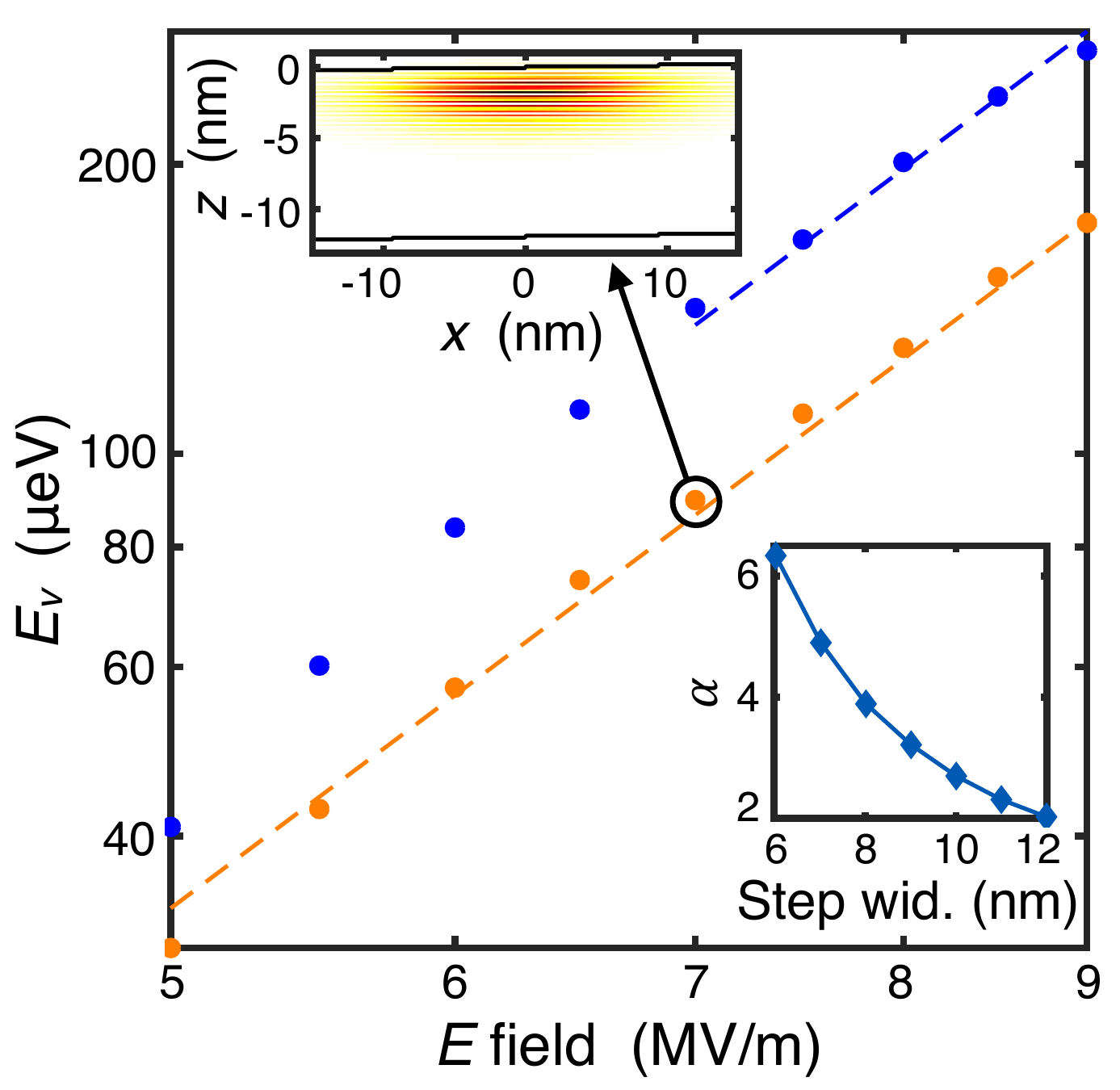}
\caption{\textbf{Tight-binding simulations of a Si quantum well with interfacial steps.} 
The upper inset shows a typical tight-binding ground-state wavefunction, in the presence of interfacial steps, for the heterostructure parameters of sample A (see supplementary material for details).
The main plot shows the valley splitting $E_v$ as a function of electric field, for filling fractions $\nu$=3 (orange circles) and $\nu$=5 (blue circles).
$E_v$ is found to be a strong function of step width; here we choose 9.4~nm, to give the best match to the experimental results in Fig.~3.
With no additional tuning parameters, we then fit both data sets to a single power law, $E_v\propto E^\alpha$, over the same field range as the experimental data (dashed lines), obtaining $\alpha$=2.8.
The correspondence with the experimental estimate ($\alpha$=2.7) is remarkable, given the strong dependence of $\alpha$ on step width, as shown in the lower inset.
}
\label{fig4}
\end{figure}

Mobility gaps corresponding to valley excitations are reported in the lower two panels of Fig.~\ref{fig3}(a), which show the extracted gaps for $\nu$=5 and 3.  While the largest gap occurs for sample C, which has enhanced Ge concentration at the top of the well, the presence or absence of such a single Ge layer does not have a dramatic effect on the valley splitting gaps we measure here, indicating that alloy disorder does not play an important role in determining the valley splitting in these samples.  In fact, all three samples reveal energy gaps that increase quite similarly with increasing perpendicular magnetic field.  One reason for this dependence is that larger magnetic fields cause electrons to occupy smaller orbits, thus mitigating the suppression of valley splitting due to interface disorder~\cite{Friesen:2006p202106,Lee:2006p245302,Goswami:2007p41}.
Valley splittings also depend strongly on the vertical electric field and thus on the density $n$.
Because the experiments are performed at two different but fixed filling fractions $\nu$, as the magnetic field changes so does the density, with $n$=$\nu eB/h$, as shown in the upper two panels of Fig.~\ref{fig3}(a). 
Large electric fields, given by $E$=$en/\varepsilon$, pull electrons strongly against the upper barrier of the quantum well, so that larger density yields larger valley splitting; for ideal interfaces with no atomic steps, the dependence of $E_v$ on $E$ is linear~\cite{Ohkawa:1977p907,Boykin:2004p115}.

A plot of the experimental results as a function of density, shown in Fig.~\ref{fig3}(b), reveals a clearly superlinear dependence on density.  Fitting simultaneously all six of the data sets (samples A-C, with $\nu$=3 and 5) to a power-law function $c_{i}n^{\alpha}$, where $\alpha$ is the same for all data sets, but $c_{i}$ is allowed to vary, yields $\alpha$=2.7$\pm$0.2. 
While the valley splitting is numerically different in all samples, all the data are fit by the same power law, as demonstrated in  Fig.~\ref{fig3}(c), where we plot the ratio of $E_v(n)/E_v^{fit}(n_0)$, with $n_0$=$4 \times 10^{11}$~cm$^{-2}$, for all data sets.

We now argue that the strong dependence of mobility gap on electric field can be understood as a consequence of steps at the quantum well interface.
We perform tight binding calculations that include the vertical electric field and interfacial  roughness, the latter in the form of uniformly spaced single-atomic steps (see supplementary material for details of the simulations).  Fig.~4(a) shows the valley splitting $E_v$ as a function of the vertical electric field for step separations of 9.4 nm. This value was chosen so that the range of valley splittings, from 30 to 200~$\mu$eV, matches the experimental measurements reported in Fig.~3.  The power law dependence of the calculated valley splitting on electric field is found to be $\alpha=2.8$, extremely close to the experimental result of 2.7.  This correspondence is remarkable, as shown in the lower right inset, which plots $\alpha$ as a function of step width, revealing that even a relatively small change in step width can easily change the power law scaling away from that shown in the main panel of Fig.~4.  
The large observed value of $\alpha$ is also remarkable for deviating so strikingly from the expectation that $\alpha$=1 \cite{Ando:1982p437,Friesen:2007p115318,Boykin:2004p115,Yang:2013p3069,Gamble:2016p253101}, which only occurs in the limit of very low disorder, as indicated by the asymptotic behavior of Fig.~4, lower inset, and supplementary material.
The similarity of the exponent $\alpha$ for the three samples is evidence that the step densities are essentially inherited from the relaxed buffer growth and underlying substrate, and do not depend on the details of the top interface or the alloy disorder occurring there. This step separation corresponds to a miscut angle ($\theta$$\simeq$0.8$^\circ$), which is larger than the sample miscut angles ($\theta$=0.1-0.2$^{\circ}$) measured with X-ray diffraction, a fact that is unsurprising, because the epitaxial growth process is expected to yield additional steps that go up and down away from the average slope. Such increases in roughness are well known in strained epitaxial growth~\cite{Evans:2012p5217}.  

Extrapolating these quantum Hall results to quantum dots is not unreasonable, with the following important caveats.
First, the energy gaps obtained by activation measurements in quantum Hall experiments are actually mobility gaps, which are affected by electron-electron interactions and localized impurities~\cite{Prange1990}.
Our estimates for the $g$-factor indicate differences between the measured mobility gap and the expected single-particle Zeeman splitting on the order of 10\%. 
Second, the quantum Hall requirement that $E\propto B$, for constant filling factor, does not apply to dots, where the confinement potential is typically defined by a fixed gate arrangement and the voltages applied to those gates.  
For example, a typical orbital energy of $\hbar\omega$=0.5~meV in a quantum dot corresponds to an r.m.s.\ radius of 20~nm, while the magnetic confinement in the $\nu$=3 Landau level at $B$=5.5~T corresponds to an r.m.s.\ radius of 7.7~nm. 
Quantum Hall transport measurements are therefore exposed to fewer atomic steps at the quantum well interface, and should typically reveal valley splittings larger than in quantum dots, for the same electric field.
Finally, it is important to note that transport measurements effectively average over mesoscopic length scales, while quantum dot measurements do not.
However, our theoretical analysis of the $\alpha$ parameter demonstrates that single-electron physics provides key insights into the observed behavior.

In summary, we conclude that it is possible to control composition in the growth direction on the very short length scales appropriate for engineering enhancements in the valley splitting \cite{Zhang:2013p2396}.
In principle, this could be a useful tool for eliminating valley splitting effects arising from alloy disorder in SiGe barriers;
however the dominant effect on the valley splitting, for the samples considered here, appears to arise from interfacial steps and atomic-scale disorder in the heterostructure layers below the top quantum well interface.
Better control of this disorder is therefore essential for increasing the valley splitting in Si/SiGe heterostructures in future experiments.

See supplementary material for details on the activation energy analysis and the tight-binding methods.

We acknowledge helpful discussions with H.-W.~Jiang, R.~Joynt, and C. A. Richter.  This research was sponsored in part by the Army Research Office (ARO) under Grant Numbers W911NF-17-1-0274, W911NF-12-1-0607 and by NSF (DMR-1206915), and by the Vannevar Bush Faculty Fellowship program sponsored by the Basic Research Office of the Assistant Secretary of Defense for Research and Engineering and funded by the Office of Naval Research through Grant No.\ N00014-15-1-0029.. Development and maintenance of the growth facilities used for fabricating samples is supported by DOE (DE-FG02-03ER46028).  The views and conclusions contained in this document are those of the authors and should not be interpreted as representing the official policies, either expressed or implied, of the Army Research Office (ARO), or the U.S. Government. The U.S. Government is authorized to reproduce and distribute reprints for Government purposes notwithstanding any copyright notation herein.  We acknowledge the use of facilities supported by NSF through the UW-Madison MRSEC (DMR-1121288).

\beginsupplement

\section*{Supplementary Information}

In these Supplementary Materials, we provide additional experimental details, and describe our theoretical models and methods.

\section{Spin Excitations at the $\bm{\nu=6}$ Landau Level}

\begin{figure}[t]
\includegraphics[width=0.3\textwidth]{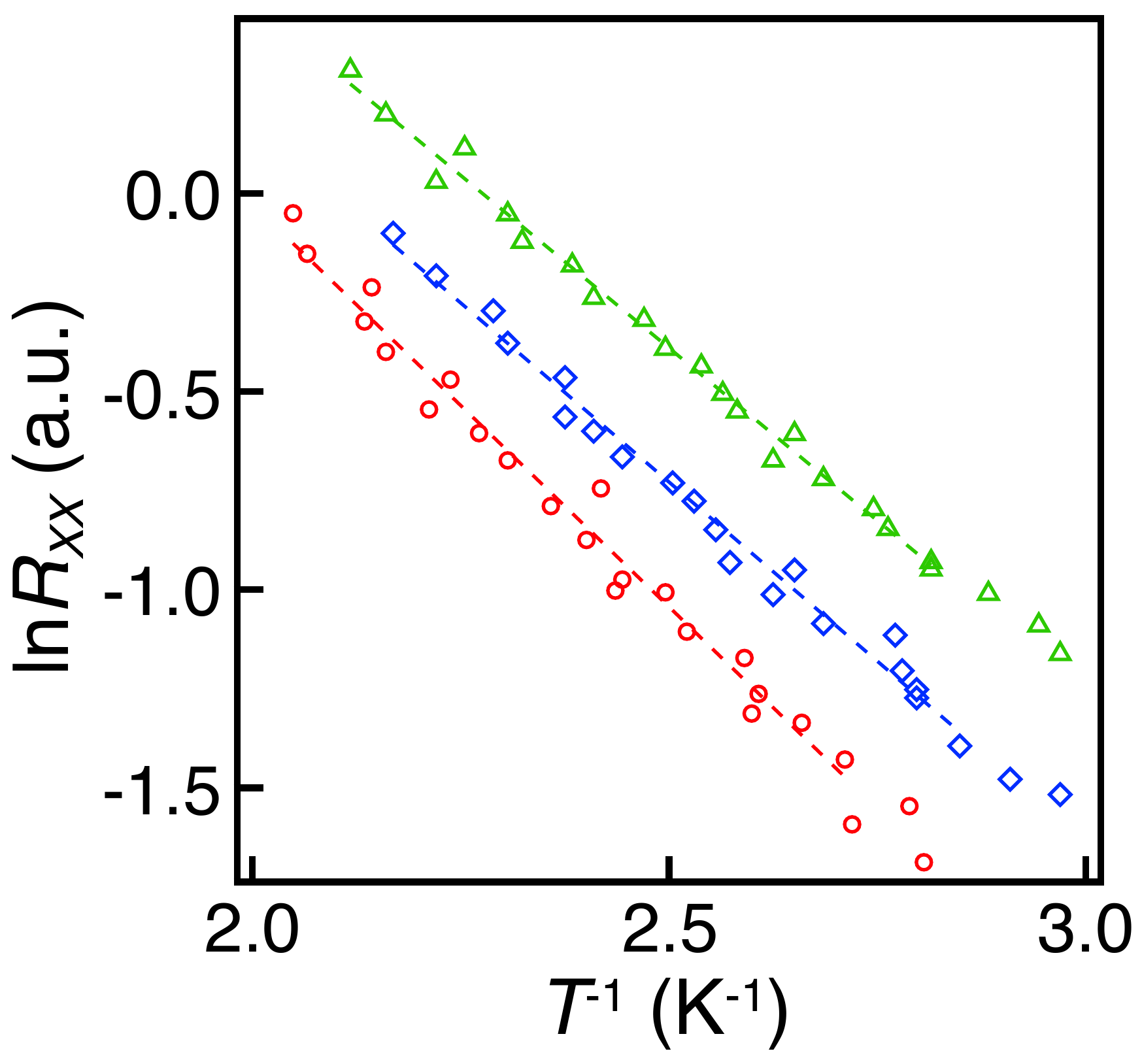}
\caption{\textbf{Thermal activation measurements of the $\bm{\nu=6}$ Zeeman splitting.} Arrhenius plots showing the dependence of the longitudinal resistance minima at $\nu=6$ on the inverse temperature, for sample A (red circles), sample B (green triangles), and sample C (blue diamonds). The data are offset vertically for clarity.  All measurements are taken at the carrier density $n=4\times10^{11}$ cm$^{-2}$. The linearity of the curves confirms the measurement is in the thermal activation regime, where $R_{XX} \propto e^{-\Delta E/2kT}$ and $\Delta E=g\mu_{B}B$ is the Zeeman splitting. The slopes of the linear fits (dashed lines) give the magnitudes of the splittings, yielding estimates for the Land\'{e} $g$-factors in the three samples. 
}
\label{fig1SM}
\end{figure}

In Fig.~\ref{fig1SM} we present the results of a spin-state activation experiment performed at $\nu=6$, which was mentioned in the main text.
The fitting procedure is analogous to the procedure discussed in Fig.~2 of the main text.
By fitting the measured activation energy to the expected Zeeman splitting, $\Delta E=g\mu_B B$, we obtain the estimates $g=2.2\pm 0.2$, $1.8\pm 0.1$, and $1.9\pm 0.2$ for the Land\'{e} $g$-factors in samples A, B, and C, respectively.
The results differ by up to 10\% from the expected value of $g\simeq 2$ for a Si quantum well.  
These variations can be attributed in part to changes in the mobility gap caused by electron-electron interactions and localization effects~\cite{Prange1990}.

\section{Tight-Binding Hamiltonian}
In order to incorporate atomic-scale disorder into our calculations, we adopt a two-band tight-binding Hamiltonian in two dimensions, as described in~\cite{Boykin:2004p115,Saraiva:2010p245314}, with nearest and next-nearest-neighbor hopping parameters in the crystallographic $\hat z$ direction, given by
\begin{equation}
v_z=4u_z\cos (k_0a/4) \quad \text{and} \quad u_z=\frac{2\hbar^2}{m_la^2\sin^2(k_0a/4)} ,
\end{equation}
where $k_0\simeq \pm 0.82(2\pi/a)$ are the $z$ valley minima in Si, for a cubic unit cell of size $a=0.543$~nm, and $m_l=0.92m_0$ is the longitudinal effective mass.
In the $\hat x$ direction, we have just one nearest-neighbor hopping parameter, given by
\begin{equation}
v_x=-\frac{8\hbar^2}{m_ta^2} ,
\end{equation}
where $m_t=0.19m_0$ is the transverse effective mass.
For the two-band model, the atomic sites are arranged on a two-dimensional square lattice with atomic spacing $a/4$.

\begin{figure}[t]
\includegraphics[width=0.47\textwidth]{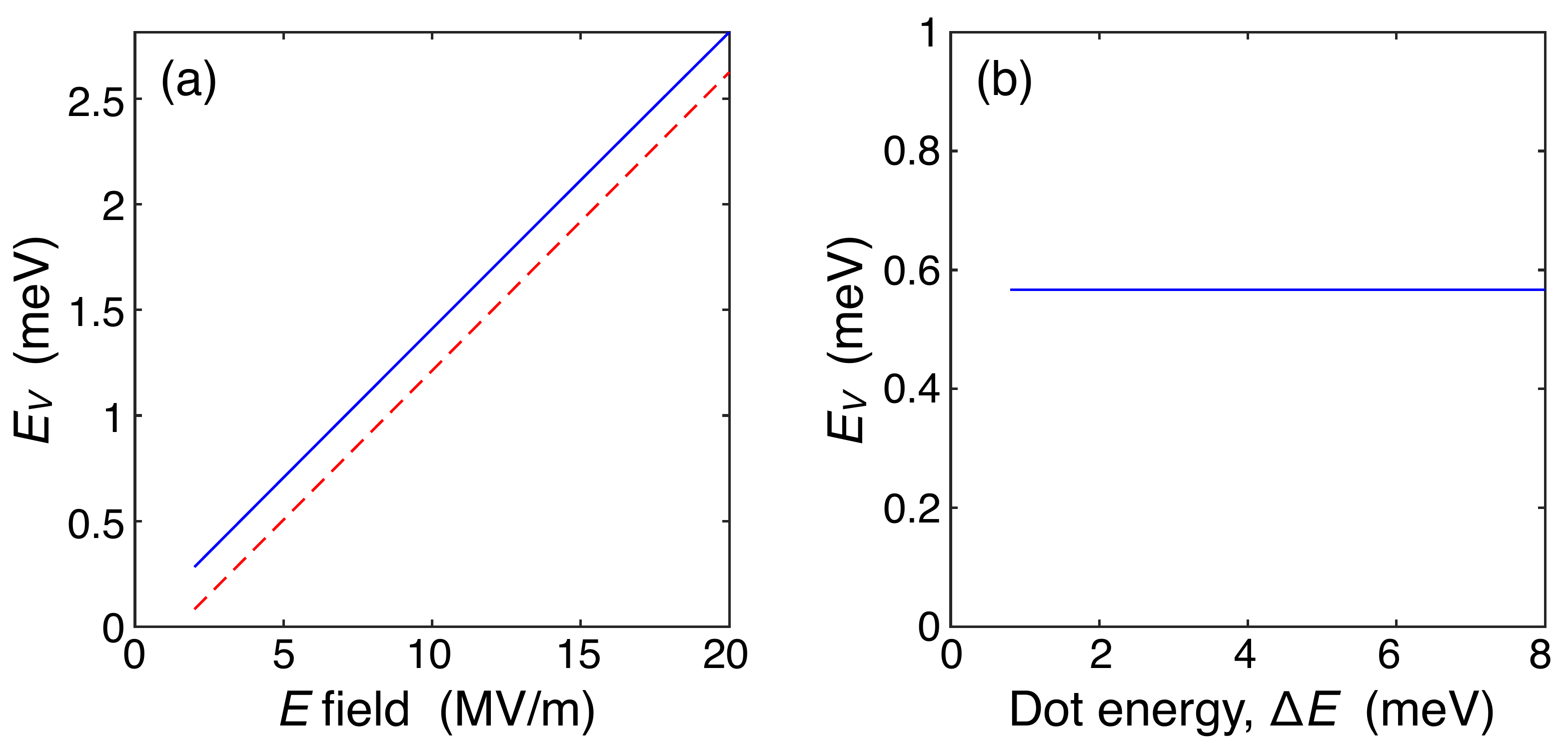}
\caption{\textbf{Tight-binding calculations of the valley splitting in a Si quantum well with no interfacial tilt, $\bm{\theta=0}$.} 
(a) Valley splitting as a function of electric field for fields in the range of 0.2-2~MV/m (blue line).
The results demonstrate a near-perfect linear dependence, as expected for a disorder-free quantum well.
(A dashed red line with slope 1 is also plotted, for reference.)
Note that the data are plotted on a linear scale, in contrast to Fig.~4 of the main text, which is plotted on a log-log scale.
For this simulation, the lateral, parabolic confinement is held fixed, assuming an orbital excitation energy of $\Delta E=\hbar \omega=4$~meV, to ensure that the two lowest eigenstates correspond to ``valley states" with the same lateral ($x'$) orbital quantum number.
(The lateral confinement potential is given by $\frac{1}{2}m_t\omega^2 {x'}^2$.)
(b) Valley splitting as a function of orbital excitation energies $\Delta E$ in the range of 0.8-8~meV.
The results indicate no dependence on $\Delta E$, as expected for a disorder-free quantum well.
Here, $E=4$~MV/m is held constant, ensuring that the two lowest eigenstates are valley states in the same orbital.
For all simulations, we assume a quantum well width of 12~nm, to ensure that the wavefunction is fully suppressed at the bottom of the well.}
\label{fig2SM}
\end{figure}

Confinement potentials describing the quantum well barrier and the magnetic and electric fields are introduced into the tight-binding Hamiltonian through the on-site terms.
For Si$_{0.71}$Ge$_{0.29}$ quantum well barriers, we assume a barrier height of 160~meV, while for Ge barriers, we assume a barrier height of 700~meV, corresponding to the $\Delta_2$ band~\cite{Schaffler:1997p1515}.
We also include atomic-scale disorder (i.e., steps) in the position of the quantum well barrier, as indicated in Fig.~4(c) of the main text.
For these simulations, we assume identical, uniformly spaced steps, which are conformal for all the different layer boundaries, yielding interfaces that are tilted away from the crystallographic $z$ axis by angle $\theta.$

The potential energy for a uniform electric field is given by $V_E=-e\mathbf{E\cdot r}$.
As in the experiments, the electric field, $\mathbf{E}=E(-\sin\theta ,0,\cos\theta )$, is oriented perpendicular to the growth surface.
As discussed in the main text, we adopt the quantum Hall constraint $n=\epsilon E/e=\nu eB/h$, which requires simultaneous variation of $E$ and $B$, to achieve a constant filling factor $\nu$.

We incorporate magnetic fields into the simulations through the Landau gauge, with the following considerations.
In the quantum Hall regime, the wave functions of current-carrying electrons are localized in one direction, but extended in the orthogonal direction, along contour lines of constant potential~\cite{Prange1990}.
For a quantum well with interfacial step disorder, such equipotential lines approximately follow the contours of the step.
To a good approximation, the magnetic confinement of an electron therefore occurs in the direction transverse to the steps, which we define here as ${\hat x}'$, relative to the growth direction, ${\hat z}'$.
For this geometry, the wavefunction component along ${\hat y}'$ is a simple plane wave, and we will ignore it here for simplicity.
The resulting magnetic confinement potential is given by
\begin{equation}
V_B=\frac{1}{2}m_t\omega_c^2(x'-x_0)^2 ,
\end{equation}
where $\omega_c=|eB/m_t|\rightarrow h\epsilon E/\nu m_te$ is the cyclotron frequency, and $x_0$ is the center of confinement for the parabolic potential.
In principle, $x_0$ could also be taken as an free parameter in our simulations; however, the valley splitting is not found to depend strongly on $x_0$, and we simply choose it to coincide with a step rise, for convenience.
To complete the tight-binding description, the potential terms $V_E$ and $V_B$ are simply evaluated at the lattice sites.

\section{Additional Tight-Binding Calculations}
In the main text, we report a superlinear dependence of the valley splitting on the carrier density $n$, and thus the electric field $E$, which we attribute to the presence of interfacial disorder.  
To justify these claims, we describe here simulation results obtained in two related model systems, in the absence of interfacial disorder.
In Fig.~\ref{fig2SM}(a), we show that the valley splitting depends linearly on $E$ for a fixed confinement potential.
In Fig.~\ref{fig2SM}(b), we show that the valley splitting is independent of the confinement potential for constant $E$.
As shown in Fig. 4 of the main text, the valley splitting depends on both $E$ and the confinement potential when interfacial steps are present.

\bibliography{main}

\end{document}